\documentclass[twocolumn,prl,showpacs,preprintnumbers,amsmath,amssymb]{revtex4}

\usepackage{graphicx}
\usepackage{subfigure}
\usepackage{dcolumn}
\usepackage{bm}

\begin{document}

\title{\textbf{\textrm{Mean-field behavior as a result of noisy local dynamics in self-organized criticality: Neuroscience implications}}}

\author{S. Amin Moosavi}
\author{Afshin Montakhab}
    \email{montakhab@shirazu.ac.ir}

\affiliation{Department of Physics, College of Sciences, Shiraz University, Shiraz 71946-84795, Iran}

\date{\today}
\begin{abstract}
Motivated by recent experiments in neuroscience which indicate that neuronal avalanches exhibit scale invariant behavior similar to self-organized critical systems, we study the role of noisy (non-conservative) local dynamics on the critical behavior of a sandpile model which can be taken to mimic the dynamics of neuronal avalanches. We find that despite the fact that noise breaks the strict local conservation required to attain criticality, our system exhibit true criticality for a wide range of noise in various dimensions, given that conservation is respected \textit{on the average}. Although the system remains critical, exhibiting finite-size scaling, the value of critical exponents change depending on the intensity of local noise. Interestingly, for sufficiently strong noise level, the critical exponents approach and saturate at their mean-field values, consistent with empirical measurements of neuronal avalanches. This is confirmed for both two and three dimensional models. However, addition of noise does not affect the exponents at the upper critical dimension ($D=4$). In addition to extensive finite-size scaling analysis of our systems, we also employ a useful time-series analysis method in order to establish true criticality of noisy systems. Finally, we discuss the implications of our work in neuroscience as well as some implications for general phenomena of criticality in non-equilibrium systems.
\end{abstract}


\maketitle

\section{Introduction}

Self-organized criticality (SOC) was proposed by Bak, Tang and Wiesenfeld (BTW) \cite{BTW1,BTW2} as a mechanism to explain generic scale invariance observed in a wide range of non-equilibrium phenomena in physical, biological and ecological systems, among others \cite{B,CM,P}. Despite its simplicity and wide range of applicability, SOC has been somewhat controversial. One objection is due to the phrase, \lq\lq self-organized\rq\rq, as it was recognized that separation of time scales -- that of driving and relaxation -- need to be implemented which implied tuning, in contrast to the original claim that no tuning was required. Thus \lq\lq slow driving" is an important feature of SOC systems \cite{foot1}. More importantly, it was quickly pointed out that conservation was the key ingredient in such systems \cite{HK1,HK2,GLS}. That is, dissipative dynamics can not lead to critical behavior. However, various non-conservative models including the Olami, Feder, Christensen (OFC) model of earthquake dynamics \cite{OFC,CO} and the Forest-Fire model \cite{DS} were proposed and studied as examples of SOC systems. Their (lack of) critical behavior has been a subject of debate in the past decades. However, recently Bonachella and Mu\~{n}oz \cite{BM} have made a strong argument for the requirement of local conservation for observation of true criticality in slowly driven non-equilibrium systems, showing that introduction of dissipation into otherwise locally conservative dynamics will introduce an effective length scale into the system and thus preventing true scale invariance in the thermodynamic limit.


In typical models of SOC, like the sandpile models, conservation is strictly obeyed on the local (microscopic) level, where the amount of energy loss at a site due to instability is exactly balanced by the amount gained by its neighboring sites ensuring local and therefore global conservation, except at the boundaries where excess energy build up due to driving is allowed to dissipate. In this paper, we propose to study a model where such restrict local conservation is broken due to the presence of noise, but \lq\lq global" conservation is respected on the average. Our main motivation is due to the recent observation of neuronal avalanches and their presumed relation to SOC. In a wide range of recent experiments \cite{BP1,BP2,FIBSLD,PTLNCP,TBFC,SACHHSCBP,HTBC,C} , neuronal avalanches have been shown to exhibit power law behavior with mean-field exponents. Whether neural dynamics \cite{ABBOTT} is dissipative or not is still debated, but their noisy dynamics \cite{RD} is a certainty, as the post-synaptic neurons receive more or less than their fair share of the ion distributed by the pre-synaptic neuron in the ionic plasma, which permeates the space between synapses. Furthermore, the question of whether violation of local conservation laws can preserve criticality if conservation holds on the average, is an interesting and important question in and of itself from a theoretical point of view.

We therefore propose to study a model of SOC with local noisy dynamics. We find that the presence of noise does not affect the criticality of the model, but modifies its critical exponents. Interestingly, we find that noise changes the critical exponents and introduces logarithmic corrections to pure power law behavior. More interestingly, we observe that for sufficiently large noise the system saturates at exponents which correspond to mean-field upper critical dimension. These are exactly the same exponents seen in the experiments on neuronal avalanches mentioned above.

The paper is constructed as follows: in Section II, we motivate and introduce our model. In Section III, we present our numerical results for a wide range of noise strengths and system sizes as well as various dimensionality while emphasizing the relevant physical issues. Finally, in Section IV, we close with some concluding remarks.

\section{The model}
In order to study the effect of noise on a SOC system, we chose the prototypical sandpile models. The discrete models, like the BTW model, are not good choices since we would like to add noise as a continuous parameter whose strength should be varied as its relative effect on the dynamics is of key interest. Besides, the BTW model exhibit peculiar scaling behavior \cite{AM}. This leads us to consider models like the Zhang model \cite{Z}. However, since we intend to study the critical behavior of our system using scaling arguments, we would like to have a model with simple finite-size scaling behavior \cite{VP}. It has been shown that a particular version of the Zhang model, known as the stochastic parallel Zhang (SPZ) model, exhibits simple finite-size scaling behavior while being free of ambiguities associated with various updating rules associated with the standard Zhang model \cite{SV}. We note that the stochasticity involved in the original SPZ model is of the same nature as that introduced in the Manna model \cite{Manna} and its many variants \cite{D,D1,MSB,Dic}. This stochasticity is introduced in the update rules in order to break the determinism in the previous (e.g. BTW) models. The \emph{noisy} dynamics we intend to study, however, breaks the conservation in the local rules.

We therefore propose to study a noisy SPZ model on a $D$-dimensional qubic lattice of size $L^{D}$.
On each site of the lattice $(j)$ we assign a continuous variable $E_{j}$ which represents the \lq\lq energy" of that site. If for all sites $E_{j}<E_{th}$, where $E_{th}=1$ is the threshold value, the system is in a stable state. The system evolves by the following dynamical rules: (i) \emph{Driving}: When the system is in a stable state some amount of energy $\delta E$ is added into a randomly chosen site $(j)$, $E_{j}\rightarrow E_{j}+\delta E$, until an unstable state is reached, i.e. the energy of one of the sites becomes larger than $E_{th}$. $\delta E$ is a random variable uniformly distributed in the range $[0,0.25]$. (ii) \emph{Toppling}: When $E_{j}$ becomes larger than $E_{th}$ the system is in an unstable state, and the site $(j)$ topples according to the rule
\begin{equation}\label{Eq.1}
\begin{array}{c}
E_{j}\rightarrow 0 \\
E_{j'}\rightarrow E_{j'}+ \epsilon_{j'} E_{j}+ \eta_{j'}
\end{array}
\end{equation}
where $(j')$ are the nearest neighbors of site $(j)$. $\epsilon_{j'}$ are annealed random numbers with the constraint $\sum\limits_{j'=n.n.} \epsilon_{j'}=1$ that ensures local conservation, in the absence of noise ($\eta=0$). $\eta_{j'}$ are random numbers chosen for every toppling from a flat noise with a zero mean value ($\langle \eta \rangle=0$) and the domain $(-\sigma,\sigma)$. We choose a flat noise for simplicity and efficiency of computer simulations. The noise violates conservation in local dynamics, however, because of its zero mean value, it respects \lq\lq global" conservation on the average. Boundaries of the system are open and  the toppling rule is applied until a stable state is reached. The sequence of these topplings which follow an initial instability is a domino like process called an avalanche. The statistical properties of such avalanches are of key interest. We therefore focus on avalanche sizes, areas and durations in order to investigate the criticality of the model. Avalanche size $(s)$ is the total number of topplings, avalanche area $(a)$ is the number of individual distinct sites that have toppled, and avalanche duration $(d)$ is the number of parallel updates, during an avalanche.

We note that this model simulates how a random neuron is activated and subsequently fires resulting in redistribution of its load to post-synaptic neurons via random ($\epsilon_{j'}$) as well as noisy ($\eta_{j'}$) synapses, thus causing other neurons to fire as well. The sum of these causal firings is considered as a single neuronal avalanche. Zero average noise is related to global charge conservation in the brain on the time scale of typical activities.

\section{Results}
We initially propose to study the criticality of our model by investigating its avalanche distribution function $P(x)$. Finite-size scaling theory of critical systems holds that such distributions obey the form $P(x)\sim x^{-\tau_{x}}f(x/L^{\beta_{x}})$ where $x$ is $s$, $d$, or $a$ depending on the quantity of interest. A system obeys simple finite-size scaling, and is therefore considered critical, if all curves for \textit{arbitrary} system size $L$, collapse into a universal form under the rescaling of $x\rightarrow x/L^{\beta_{x}}$ and $P(x)\rightarrow L^{\tau_{x} \beta_{x}}P(x)$ \cite{VP,SV}. Some systems are found to require an additional exponent through what is known as logarithmic correction in order to better fall into a single universal curve, i.e. $P(x)\sim x^{-\tau_{x}} (\ln(x))^{\gamma_{x}} f(x/L^{\beta_{x}})$  and collapse under rescaling of $x\rightarrow x/L^{\beta_{x}}$ and $P(x)\rightarrow L^{\tau_{x} \beta_{x}}(\ln(x))^{-\gamma_{x}}P(x)$. One may also attempt to collapse the data by rescaling the $y$-axis with $x^{\tau_{x}}$ instead of $L^{\tau_{x}\beta_{x}}$. We have used both methods in what follows.

\begin{figure}
\begin{center}
\subfigure[]{\includegraphics[width=0.4\textwidth]{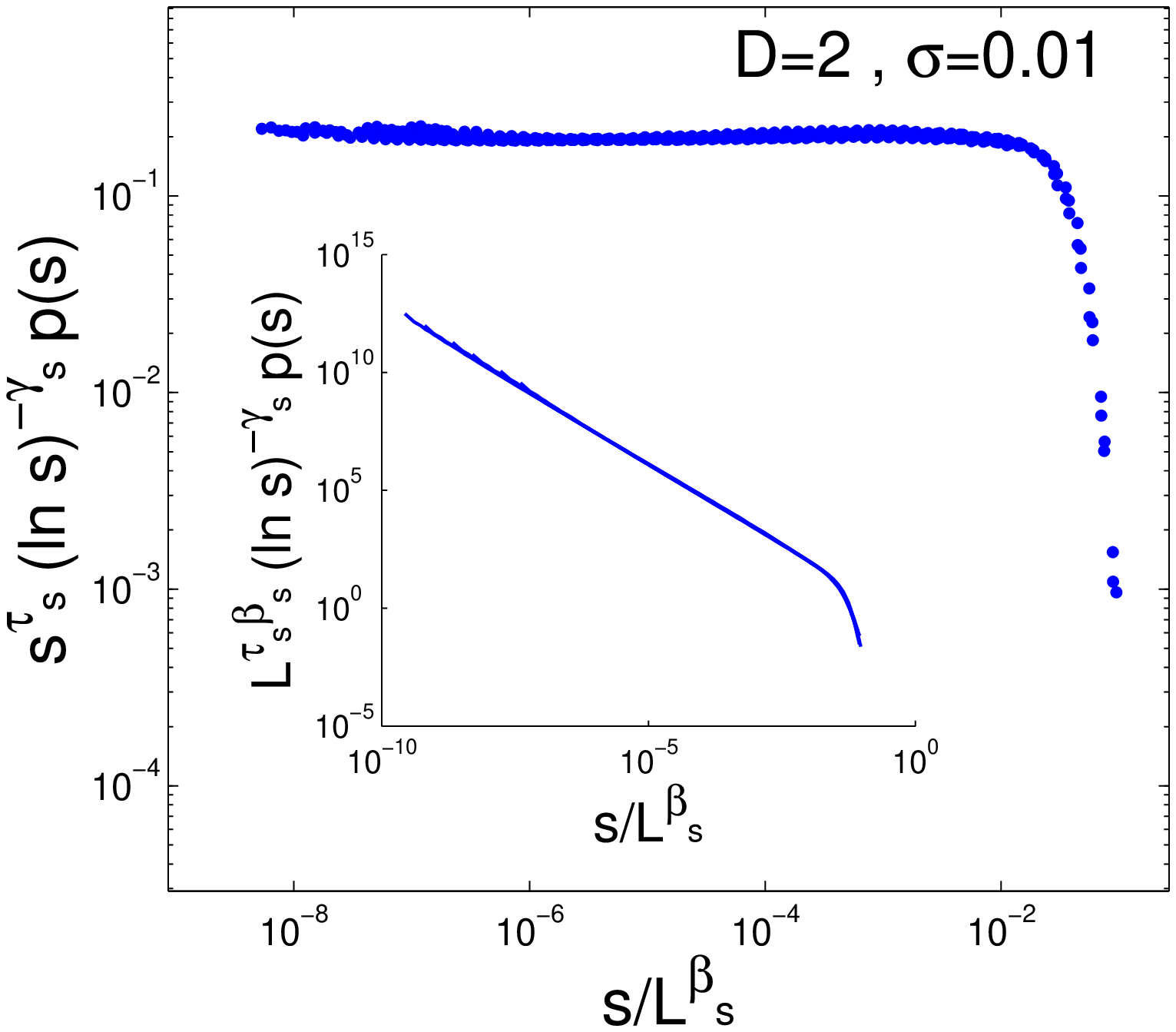}\label{fig1a}}
\subfigure[]{\includegraphics[width=0.4\textwidth]{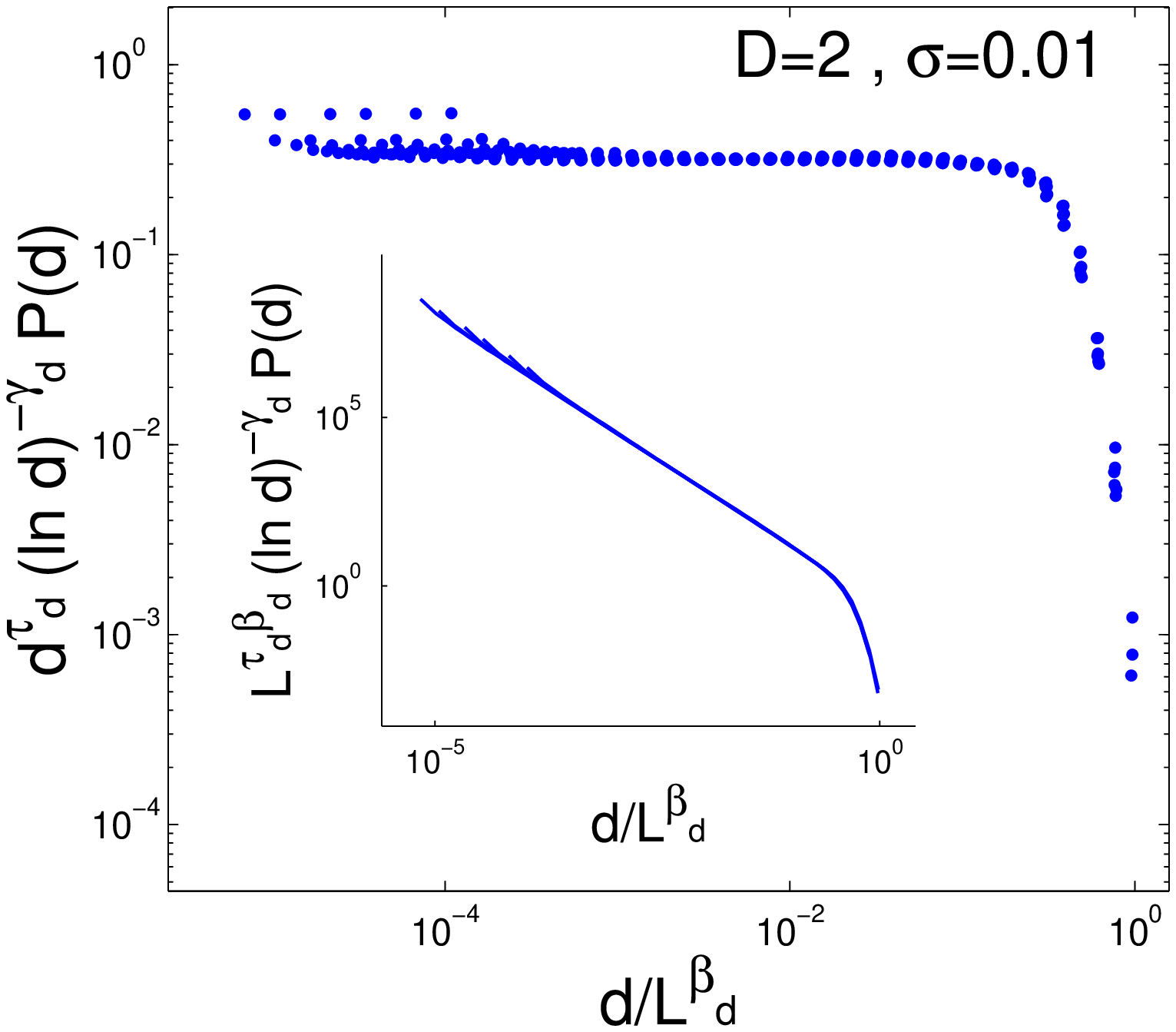}\label{fig1b}}
\subfigure[]{\includegraphics[width=0.4\textwidth]{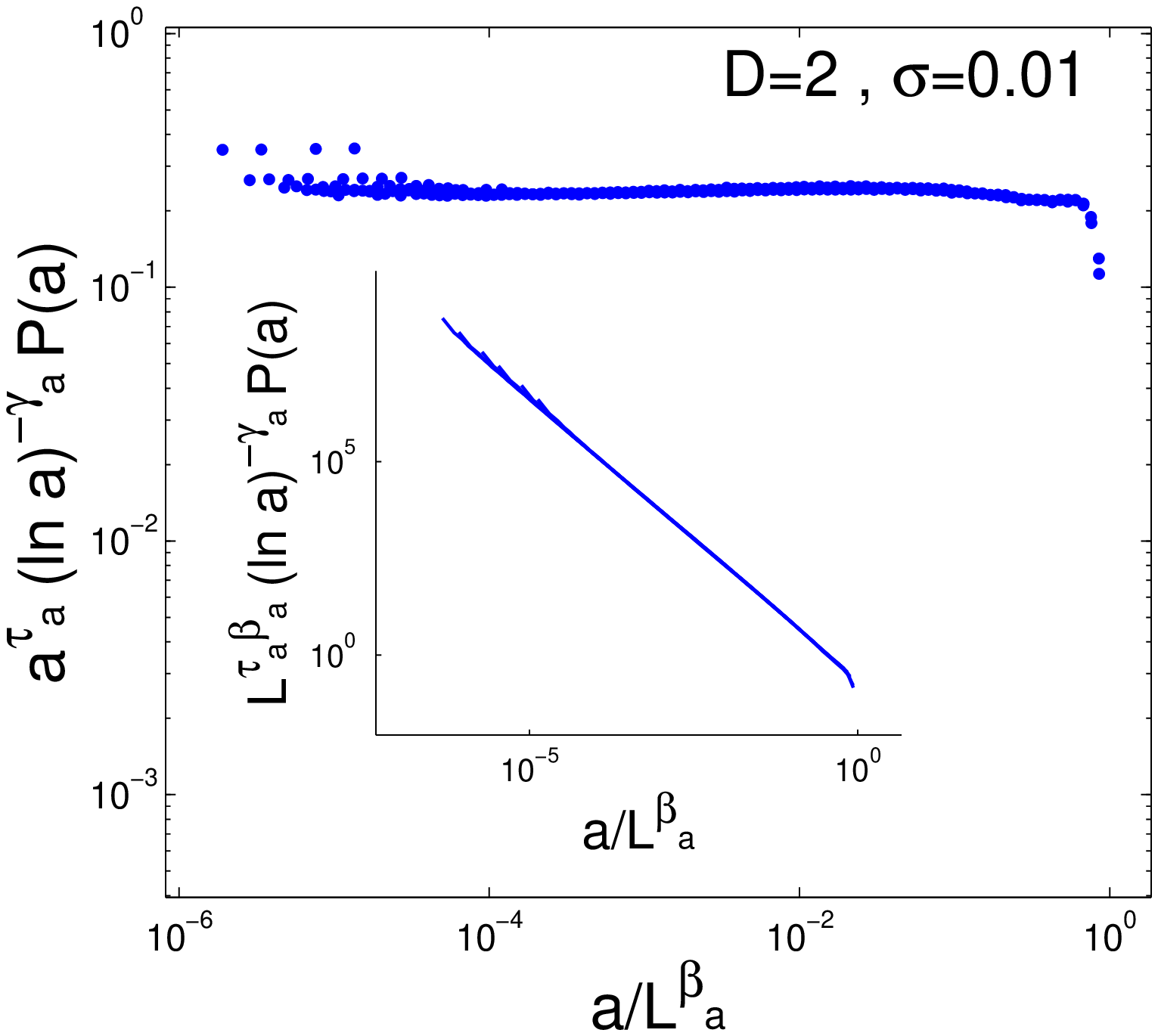}\label{fig1c}}
\end{center}
\caption{Finite-size-scaling collapse for (a) size, (b) duration and (c) area of avalanches for two dimensional SPZ model. Linear system sizes are $L=384, 512, 768, 1024, 1536, 2048$ and $\sigma=0.01$. The main panels show the collapse using the activity variable ($x$) while the insets show the same data collapsed with system size ($L$). In both methods we observe good collapses. We report the exponents obtained from the main panel in Table \ref{table1}.}
\label{fig1}
\end{figure}

\begin{figure}
\begin{center}
\subfigure[]{\includegraphics[width=0.4\textwidth]{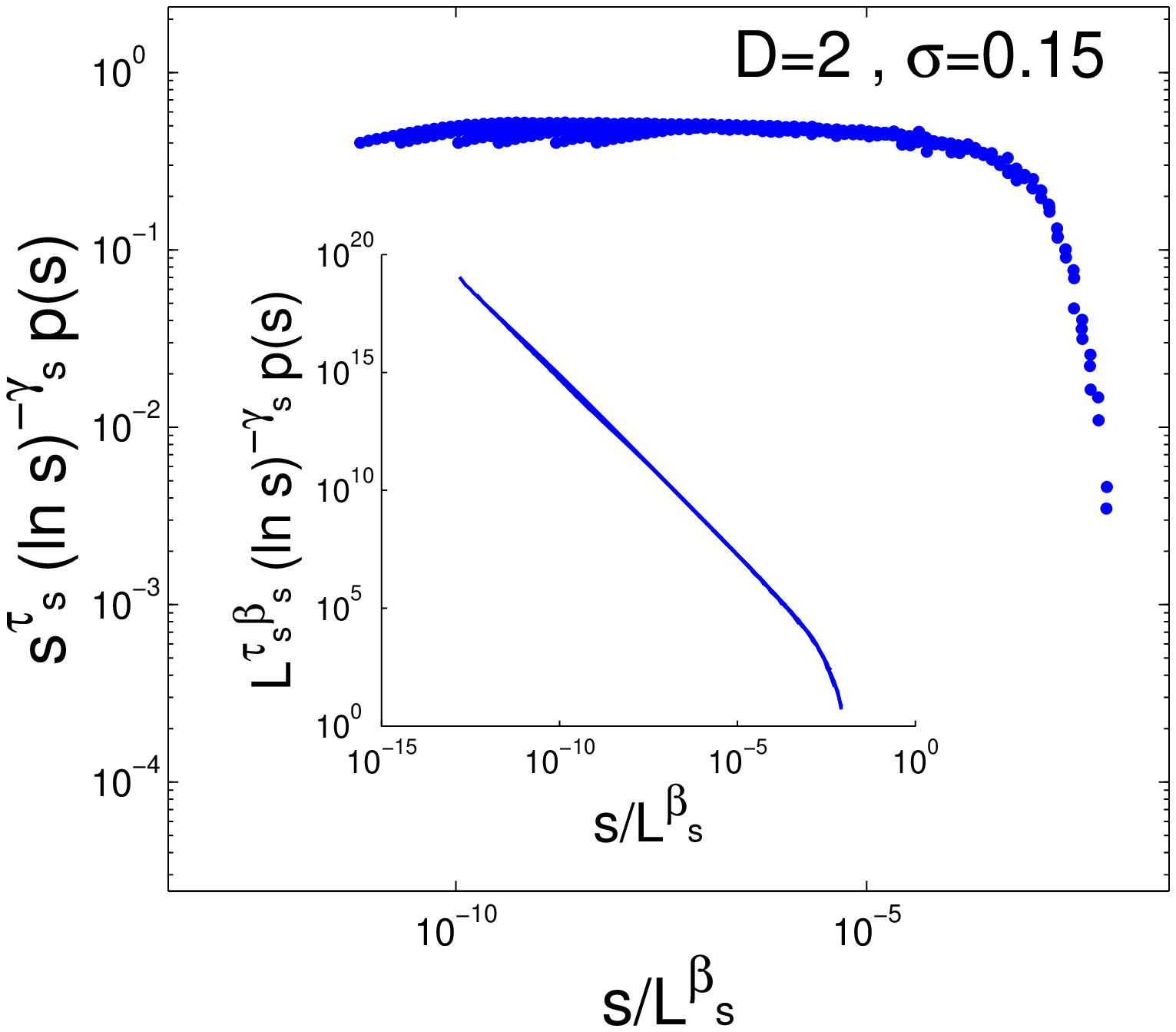}\label{fig2a}}
\subfigure[]{\includegraphics[width=0.4\textwidth]{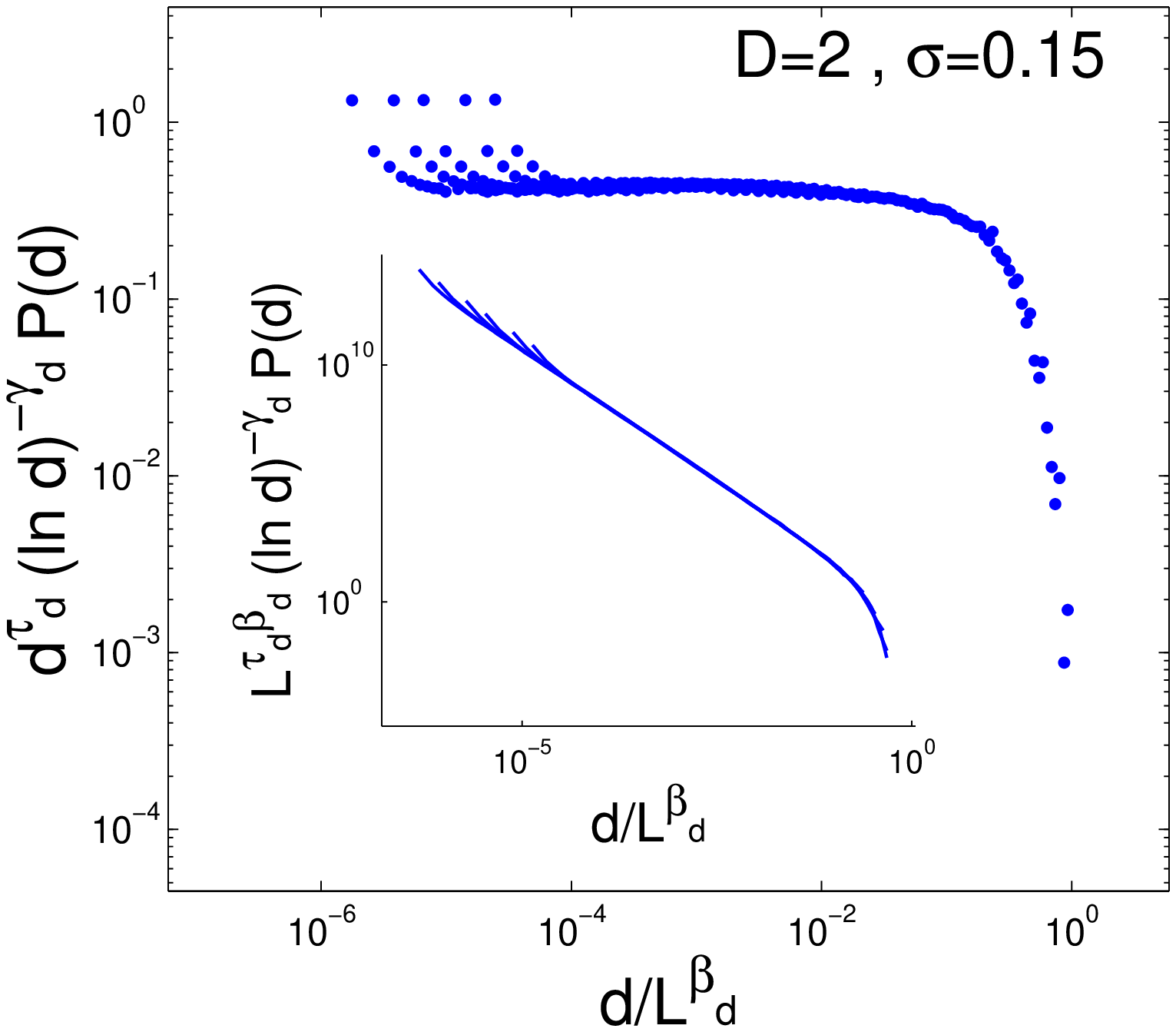}\label{fig2b}}
\subfigure[]{\includegraphics[width=0.4\textwidth]{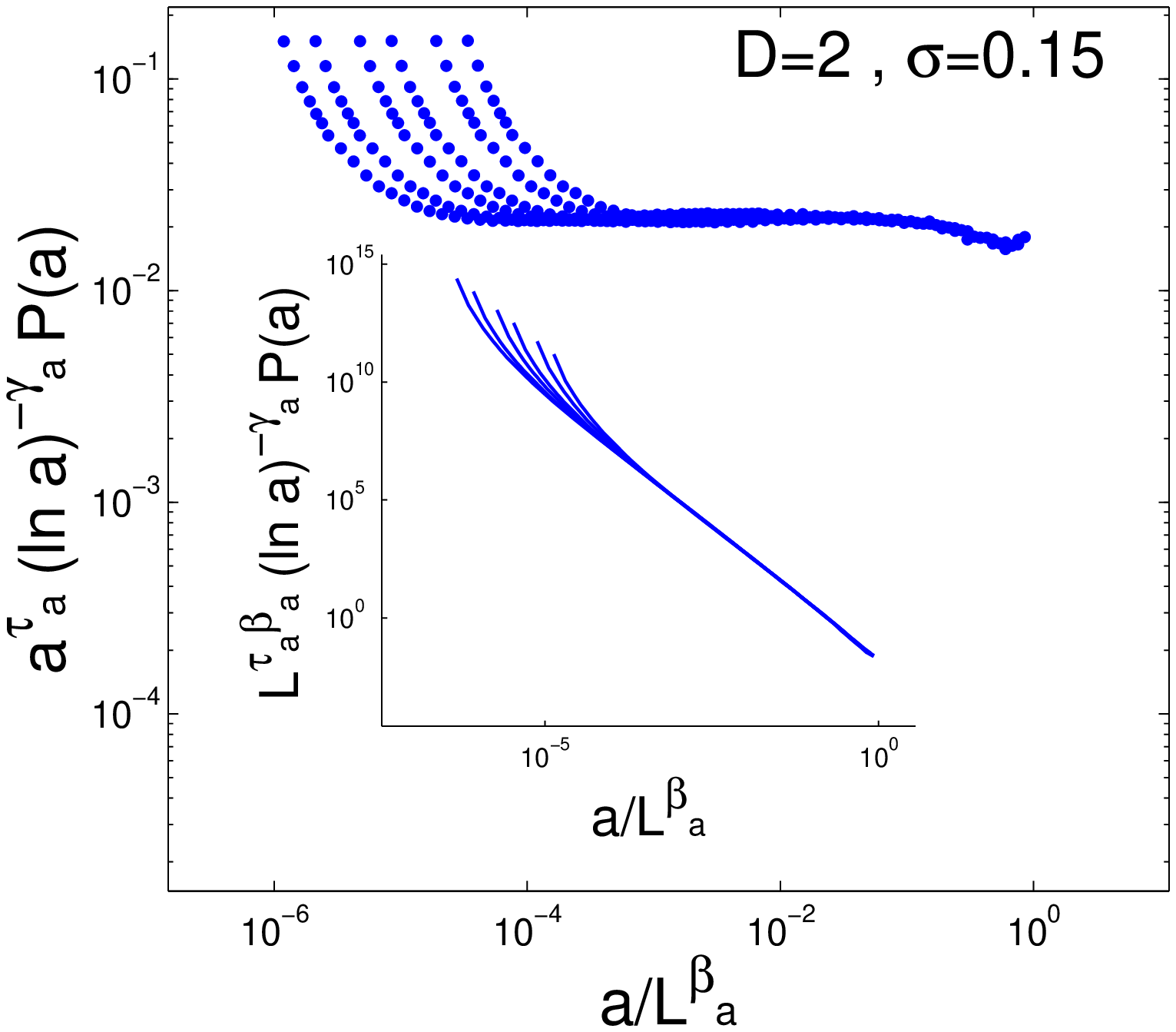}\label{fig2c}}
\end{center}
\caption{Finite-size-scaling collapse for (a) size, (b) duration and (c) area of avalanches for two dimensional SPZ model. Linear system sizes are $L=384, 512, 768, 1024, 1536, 2048$ and $\sigma=0.15$. The main panels show the collapse using the activity variable ($x$) while the insets show the same data collapsed with system size ($L$). In both methods we observe good collapses. We report the exponents obtained from the main panel in Table \ref{table1}.}
\label{fig2}
\end{figure}

We therefore have simulated our model up to system size of $L=2048$ in two dimensions creating up to $2\times10^{7}$ avalanches, and have perform finite-size scaling collapses in order to obtain exponents $\tau_{x}$, $\beta_{x}$, $\gamma_{x}$ for $x=s, a, d$. We start from restrictly conservative model $\sigma=0$ where our results match that of previous studies \cite{SV}. We then increase the level of noise and find good finite-size collapses for various noise levels $\sigma$. We note that the presence of noise increases simulation time considerably, not only due to its implementation at each time step but also because of its effect on increasing duration (and size) of a given avalanche. We present a few examples of our finite-size scaling plots in Fig.\ref{fig1} and \ref{fig2}, and report the totality of our results in Table \ref{table1}. We note that in all cases a good collapse is obtained (particularly for $s$ and $d$) indicating the validity of criticality in the presence of noise. However, another important feature of the behavior is immediately seen. As is seen in Table \ref{table1} (D=2), with increasing $\sigma$ the avalanche size and duration exponents quickly increase to about their mean-field values of $\tau_{d}=2$ and $\tau_{s}=1.5$ \cite{A,ZLS}. This increase is accompanied by a significant amount of logarithmic correction ($\gamma_{x}\neq0$). However, as noise level ($\sigma$) is further increased the critical exponents find and saturate at their exact mean-field values, where little or no logarithmic corrections is needed.

Logarithmic correction to power law scaling has been seen in various models of SOC including the two dimensional Manna model \cite{DC}, BTW model on scale free networks \cite{GLKK}, as well as BTW model at upper critical dimension ($D=4$) \cite{LU}. We note that logarithmic corrections near upper critical dimension is predicted by renormalization group theory \cite{W}. Therefore, the emergence of non-zero $\gamma_{x}$ as $\sigma$ is increased is consistent with the observed approach to mean-field (upper-critical dimension) results.

\begin{figure}
\begin{center}
\subfigure[]{\includegraphics[width=0.4\textwidth]{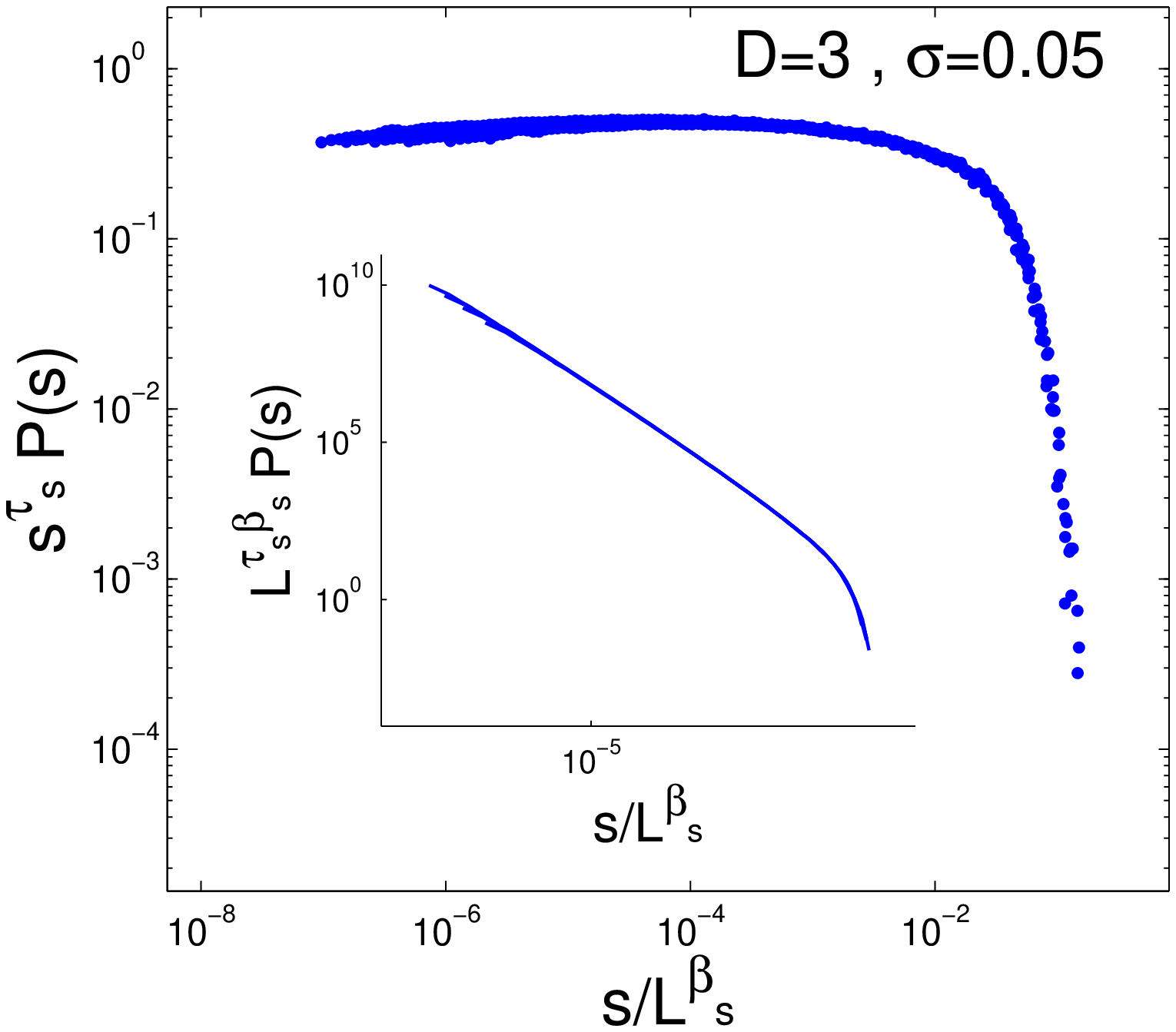}\label{fig3a}}
\subfigure[]{\includegraphics[width=0.4\textwidth]{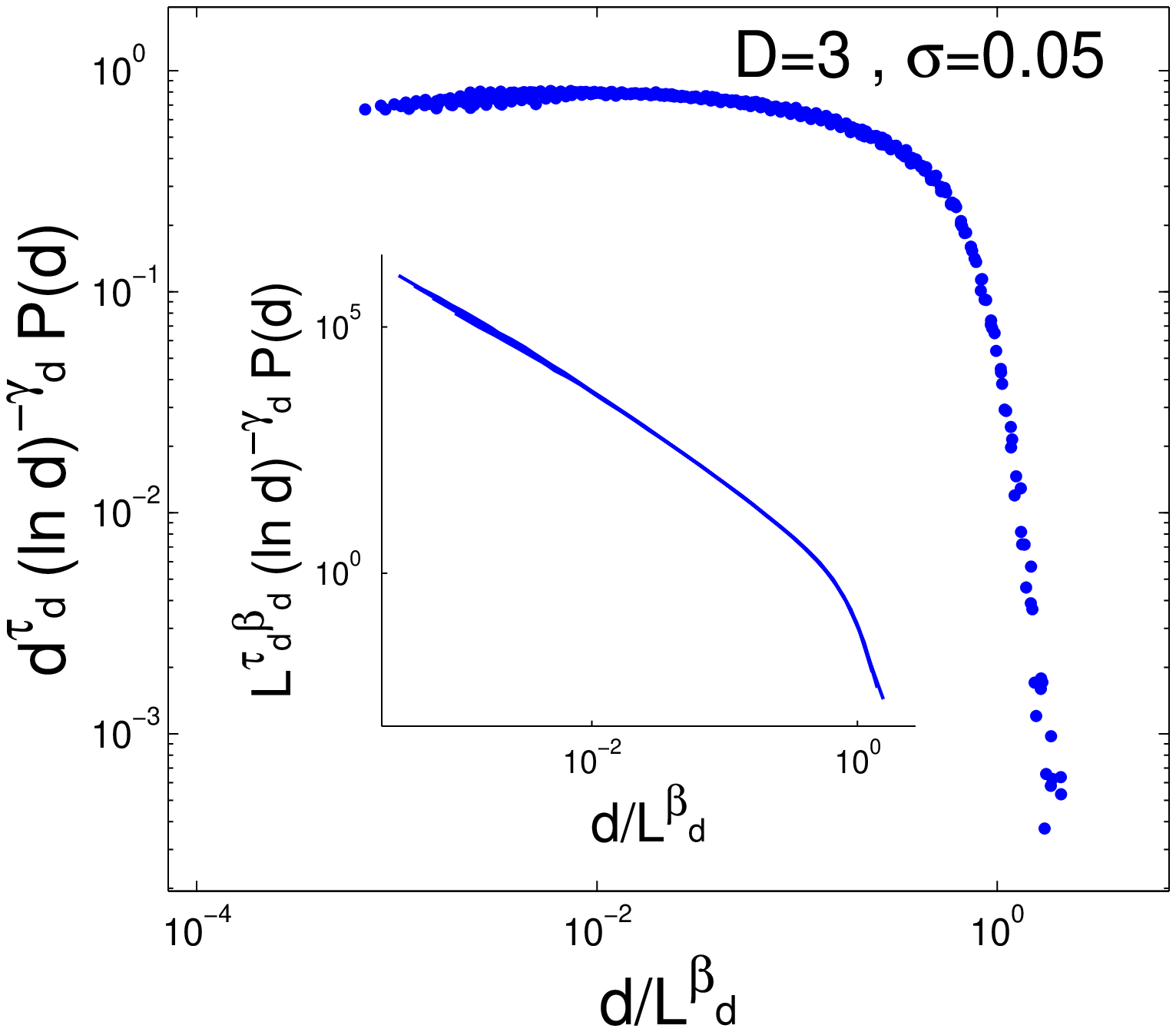}\label{fig3b}}
\subfigure[]{\includegraphics[width=0.4\textwidth]{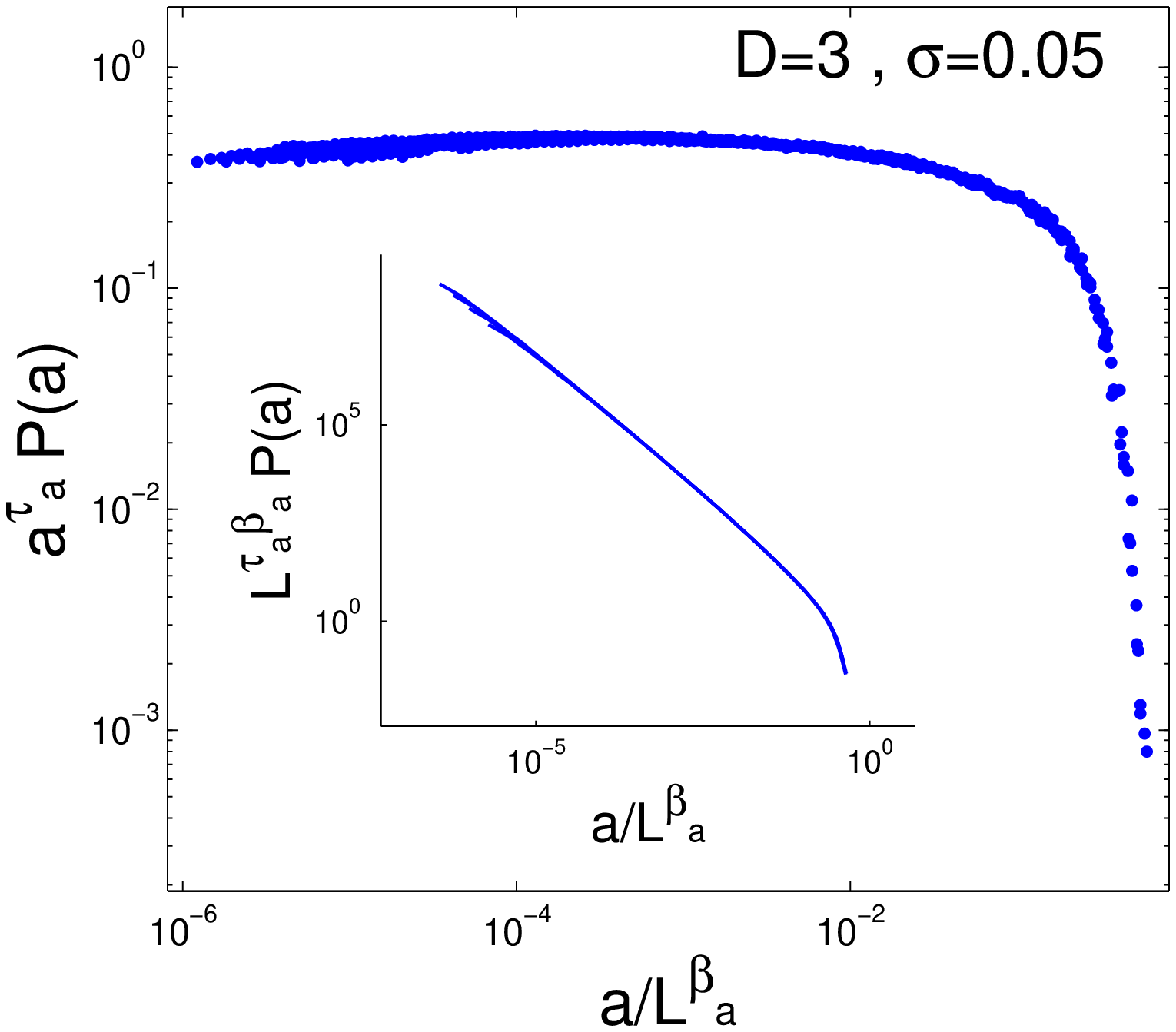}\label{fig3c}}
\end{center}
\caption{Finite-size-scaling collapse for (a) size, (b) duration and (c) area of avalanches for three dimensional SPZ model. Linear system sizes are $L=80, 100, 120, 140, 160$ and $\sigma=0.05$. The main panels show the collapse using the activity variable ($x$) while the insets show the same data collapsed with system size ($L$). In both methods we observe good collapses. We report the exponents obtained from the main panel in Table \ref{table1}.}
\label{fig3}
\end{figure}

\begin{figure}
\begin{center}
\subfigure[]{\includegraphics[width=0.4\textwidth]{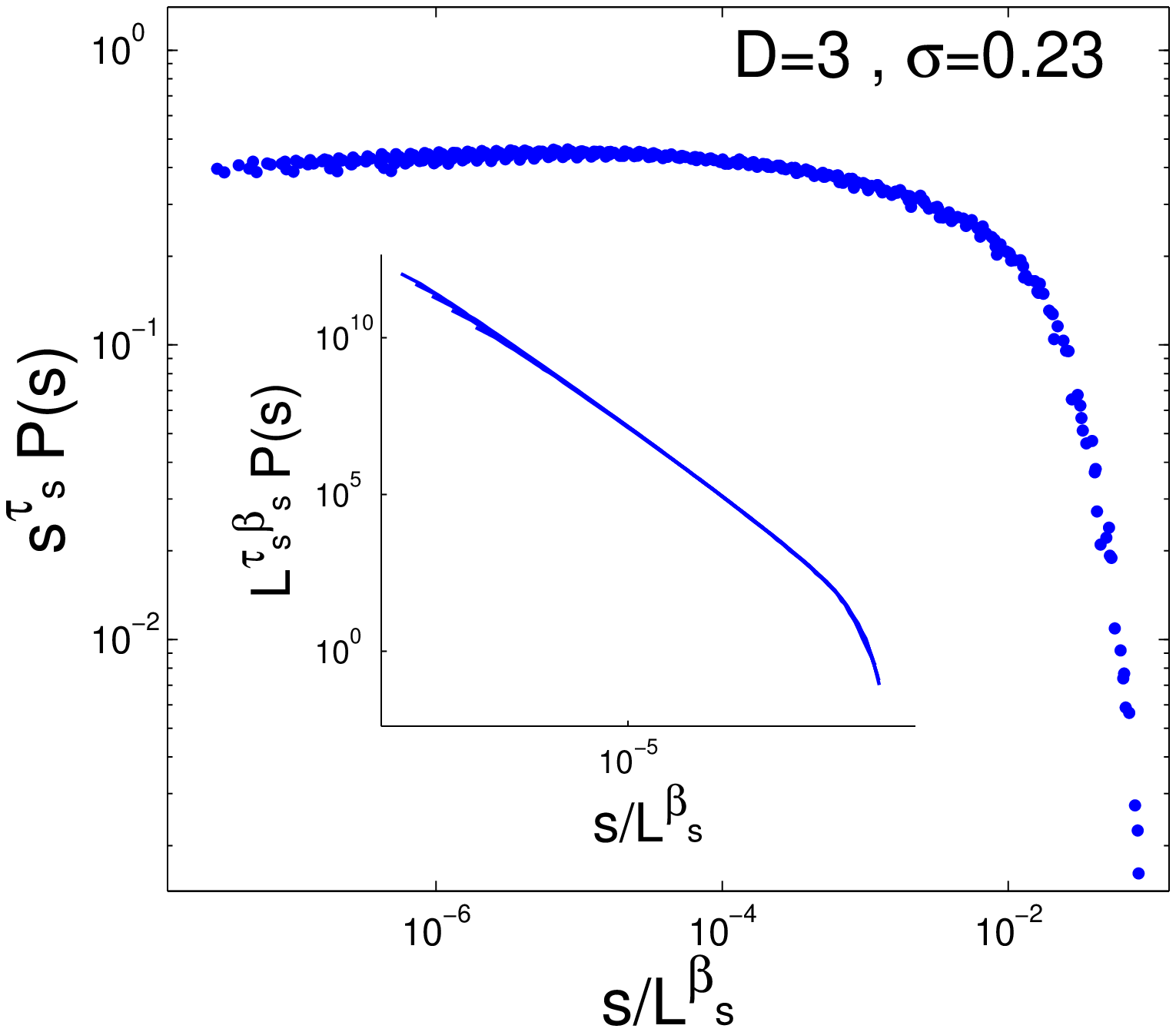}\label{fig4a}}
\subfigure[]{\includegraphics[width=0.4\textwidth]{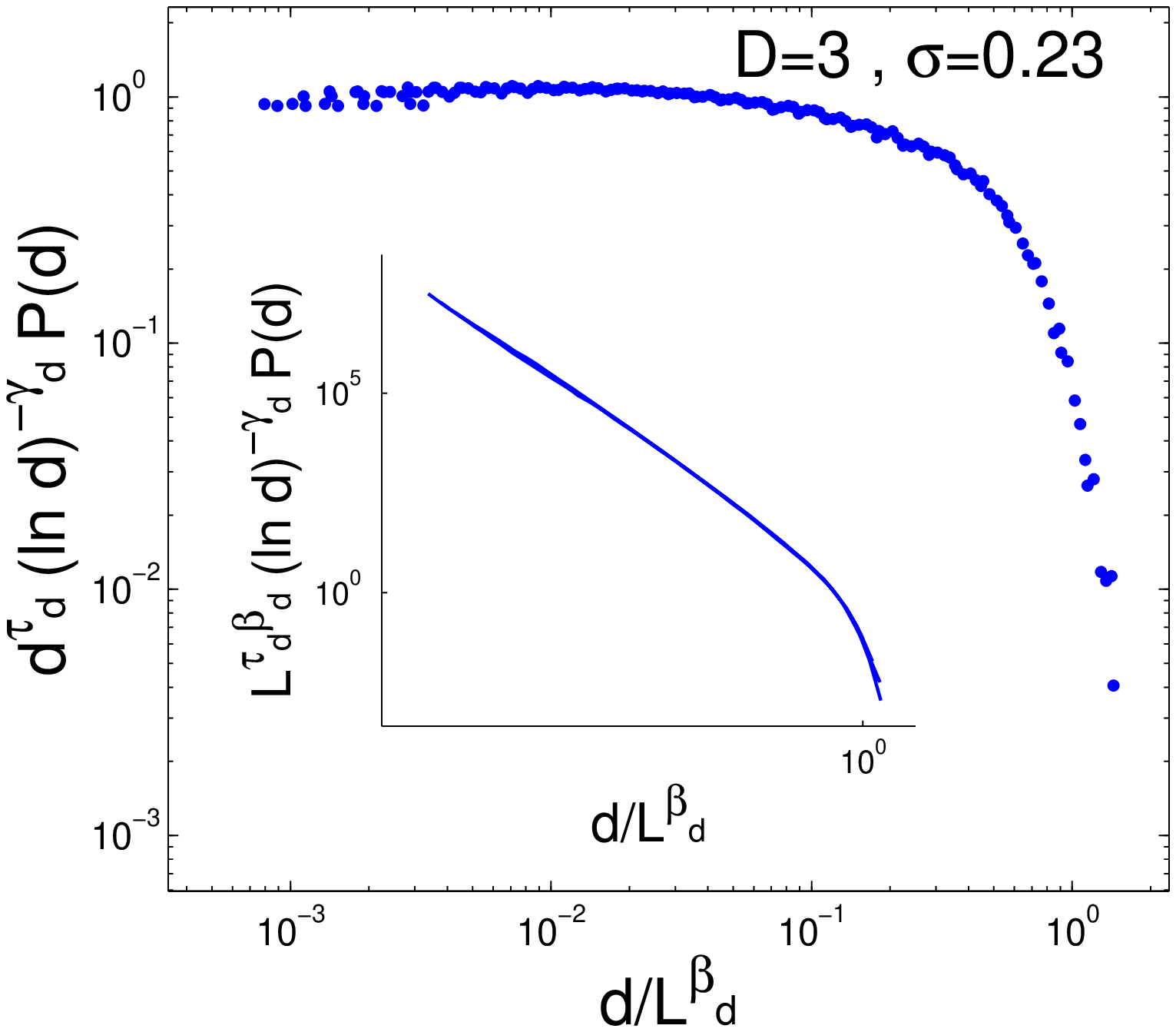}\label{fig4b}}
\subfigure[]{\includegraphics[width=0.4\textwidth]{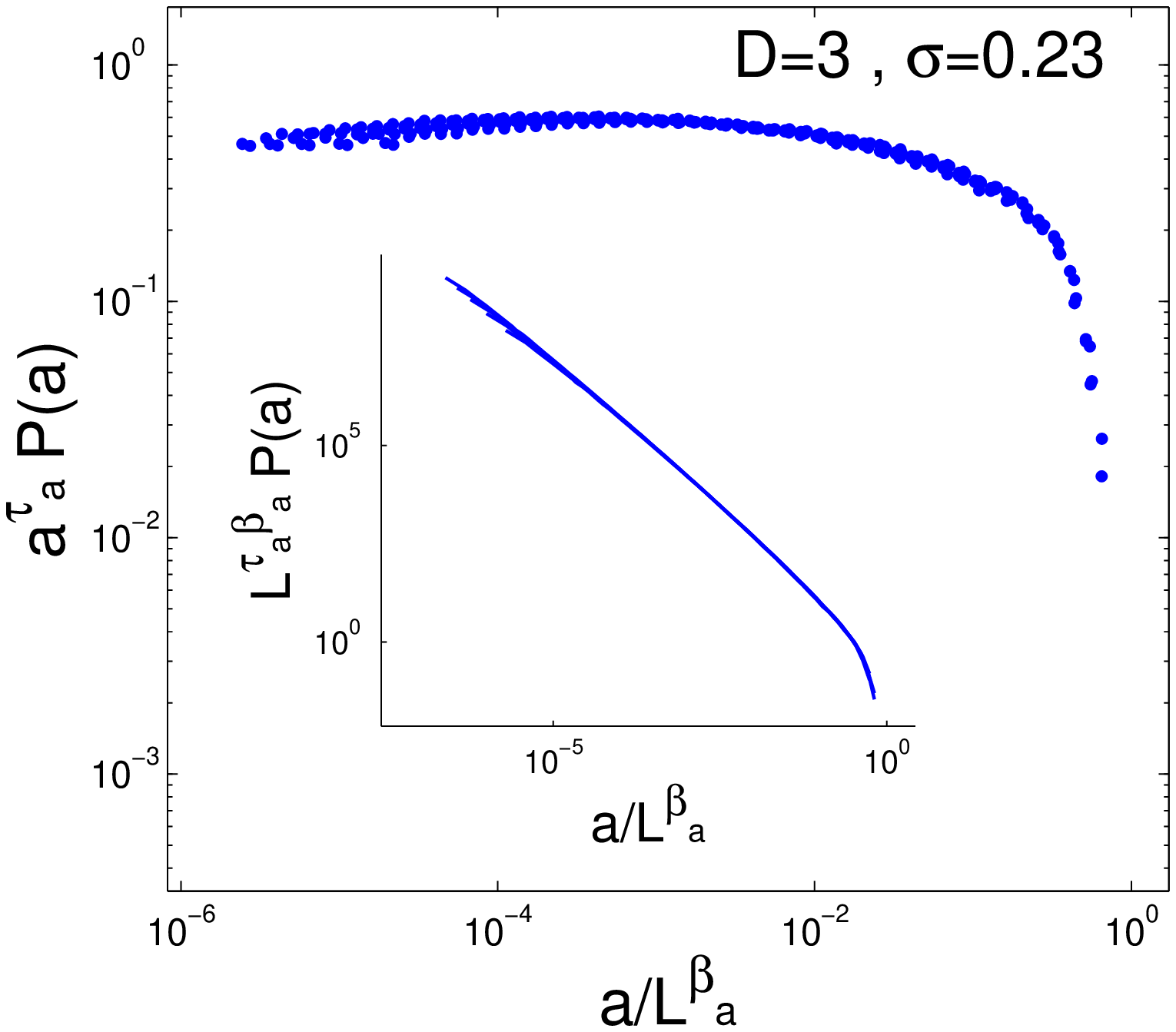}\label{fig4c}}
\end{center}
\caption{Finite-size-scaling collapse for (a) size, (b) duration and (c) area of avalanches for three dimensional SPZ model. Linear system sizes are $L=80, 100, 120, 140, 160$ and $\sigma=0.23$. The main panels show the collapse using the activity variable ($x$) while the insets show the same data collapsed with system size ($L$). In both methods we observe good collapses. We report the exponents obtained from the main panel in Table \ref{table1}.}
\label{fig4}
\end{figure}

In order to further investigate whether the $2D$ noisy SPZ model is at or near upper critical dimension, we have performed three and four dimensional simulations of the SPZ model with and without noise, where we have plotted a sample of our finite-size scaling collapses in Fig.\ref{fig3} and \ref{fig4} for $D=3$. Note that the distinction between avalanche size and area disappears in a mean-field treatment \cite{A,ZLS}. Accordingly, we observe little difference in $D=3$ (Table \ref{table1}) and no difference at all in $D=4$ (Table \ref{table2}) between $s$ and $a$ in our simulations. One can see that $D=4$ gives mean-field exponents, i.e. $\tau_{d}=2.0$ and $\tau_{s}=1.5$ along with $\gamma_{s}=0.00$ and $\gamma_{d}=0.80$. More importantly, we note that in $D=3$ (Table \ref{table1}) the rise in critical exponents with increasing noise is more gradual as they reach their mean-field values. Interestingly, there is no logarithmic corrections needed in $D=3$ for various noise levels beyond what is needed at the upper critical dimension $D=4$ (see Table \ref{table2}).  We also note that our main focus here is on exponents $\tau_{s}$ and $\tau_{d}$ as they are the exponents measured in typical (neuronal avalanche) experiments. They are also the exponents which are directly obtained in the mean-field solutions of sandpile models \cite{A,ZLS}. The $\beta$ exponents are clearly finite-size exponents and not treated as critical exponents of the system, and therefore one would not expect to compare them for various dimensionality and models. As reported in Table \ref{table2}, the noiseless (conserved) $4D$ model exhibits mean-field exponents $\tau_{d}$,$\gamma_{d}$,$\tau_{s}$,$\gamma_{s}$ which are identical to $2D$ and $3D$ systems at $\sigma=0.23$. We also note that addition of noise does not affect the critical exponents of $4D$ model, as shown in Table \ref{table2} for $\sigma=0.23$. We emphasize that the critical exponents ($\tau_{s},\gamma_{s},\tau_{d},\gamma_{d}$) gradually rise and reach their mean-field values as a function of $\sigma$ in $3D$ while in $2D$ they initially overshoot with significant logarithmic correction before finding and settling at their mean-field values.

\begin{figure}
\begin{center}
\includegraphics[width=8cm,height=7cm]{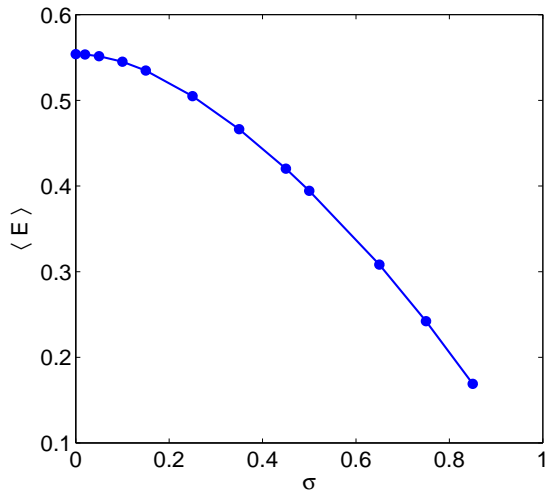}
\end{center}
\caption{Average energy $\langle E\rangle$ versus noise intensity $\sigma$ for a two dimensional noisy system. Increasing noise reduces the average energy significantly.}
\label{fig5}
\end{figure}

In addition to critical exponents, we have also looked at the \lq\lq critical point" of the various models. The critical point of such systems are obtained as the time average of the mean total energy ($\langle E\rangle=\langle\sum_{i}E_{i}/L^{D}\rangle$) where a balance between drive and dissipation at the boundaries is reached, in the thermodynamic limit $L\rightarrow \infty$ \cite{MC}. In the absence of noise, increasing $D$ reduces the average energy $\langle E\rangle$ as system finds more boundaries for energy dissipation. Interestingly, as can be seen from Table \ref{table1} increasing noise has the same effect. Fig.\ref{fig5} shows the average energy as a function of noise level for $D=2$. Significant decrease in the average energy is seen for large values of noise. We note that the $2D$ system with $\sigma=0.23$ exhibit not only the same critical exponents as the $4D$, it also exhibit the same critical point $\langle E\rangle = 0.512$ (see Table \ref{table2}). However, we believe this is just a coincidence as it is not born out by $3D$ results $\langle E\rangle^{\sigma=0.23}_{3D}=0.478$. We also point that while obtaining the same critical exponents is of physical significance (e.g. universal behavior) not much meaning can be attached to the actual values of critical points.

\begin{figure}
\begin{center}
\includegraphics[width=8cm,height=7cm]{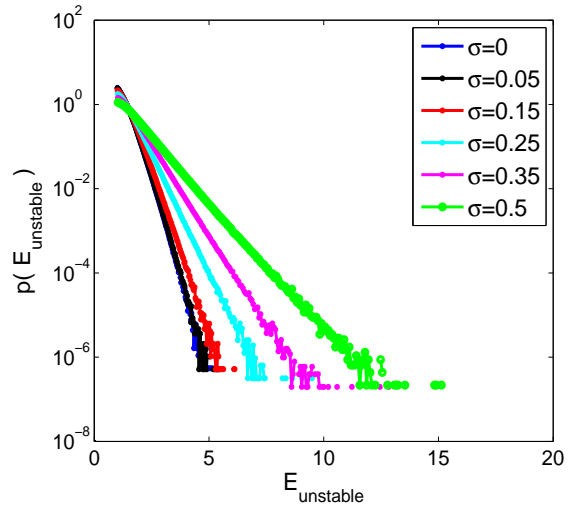}
\end{center}
\caption{(Color online) Probability density function (frequency of occurrence) of the energy of unstable sites just before toppling for different values of noise intensity $\sigma$, on a two dimensional system of $L=1024$. }
\label{fig6}
\end{figure}

But why should the addition of noise lead to mean-field results? Increasing dimension of the system ($D$) reduces average energy $\langle E\rangle$ while increases the number of nearest neighbors ($=2\times D$ on a cubic lattice) through which an unstable site can directly influence. Well, increasing noise has exactly the same effect: First of all, noise with average of zero is only possible with negative values which consequently can lead to negative values of $E_{j}$ thus effectively reducing the average energy much like increasing dimension does (see Fig.\ref{fig5}). Secondly, and on the other hand, large possible values of noise can lead to values of unstable sites which are much larger than threshold, i.e. $E\gg E_{th}$. Fig.\ref{fig6} shows the probability distribution of the values of super-threshold sites as measured before they are toppled. Relaxation of such super-threshold sites and their subsequent transfer to their nearest neighbors leaves them well above threshold and thus effectively transports their energy on their next and next nearest neighbors. Therefore, increasing noise, much like increasing $D$ leads to more and more \lq\lq effective" neighbors, and consequently leads to a mean-field behavior.

\begin{figure}
\begin{center}
\includegraphics[width=8cm,height=7cm]{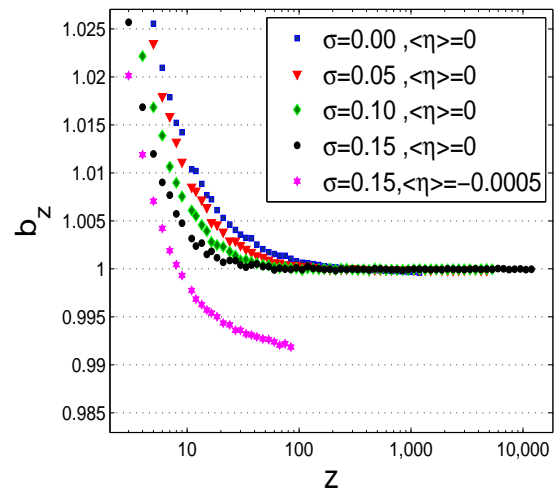}
\end{center}
\caption{(Color online) Activity dependent branching ratio $b_{z}$ for a two dimensional noisy system of $L=1024$. All systems where $\langle\eta\rangle=0$ exhibit $b_{z}=1$ for a wide range of data indicating true criticality. In the case of $\langle \eta\rangle\lesssim 0$, it is clearly seen that $b_{z}<1$ indicating a sub-critical system.}
\label{fig7}
\end{figure}

Lastly, we discuss the issue of apparent vs. true criticality \cite{BM}. Finite-size scaling is the true indicator of a critical system. However such methods are at times prone to error as obtaining enough statistics for larger system sizes are computationally limited, and thus one is forced to relay on approximation methods \cite{foot2} which may lead to scaling for finite systems but fail in a truely large (infinite) system size, which is what is meant by the phrase \lq\lq apparent criticality". Branching ratio is a useful method of studying non-equilibrium systems. However, branching ratio is a globally defined quantity useful in mean-field studies. Recently, a method based on time series analysis of branching ratios has been proposed which assigns a branching ratio to a given activity and  is a reliable method of distinguishing criticality \cite{MSP}. Activity dependent branching ratio ($b_{z}$) is defined for a time series $\{Z_{t}\}$ as the expectation value of $\xi_{z}/z$, i.e. $b_{z}=E(\xi_{z}/z)$, in which $\xi_{z}$ is the value of $Z_{t+1}$, when $Z_{t}$ is given to be equal to $z$, i.e. $\xi_{z}=(Z_{t+1}|Z_{t}=z)$. In a true critical system $b_{z}$ is expected to be exactly one over a wide range of data, preventing any short term predictability, while one expects $b_{z}<1$ ($b_{z}>1$) for a sub (super) critical system. We have therefore produced a time series of $\{Z_{t}\}$ where $Z_{t}$ is the number of unstable sites at time $t$, and $t$ is the parallel update time, and have calculated $b_{z}$ for various two dimensional models and plotted the results in Fig.\ref{fig7}. The fact that $b_{z}=1$ for all noise levels including $\sigma=0$ is a strong indication of the criticality of the system regardless of the noise level. We note that it has been seen \cite{MSP,MM} that a smallest level of dissipation introduced into an otherwise locally conserved dynamics will lead to a sub-critical behavior which gives $b_{z}<1$ over a wide range of $z$. In our model if we introduce a non-conserving noise ($\langle\eta \rangle \lesssim 0$), we also note that $b_{z}$ is reduced as is seen in Fig.\ref{fig7} for $\langle\eta \rangle=-0.0005$. Not only $b_{z}<1$ for dissipative system, the range of data is significantly reduced indicating sub-critical behavior.

\begin{table*}

\setlength{\tabcolsep}{5pt}
\begin{tabular}{l | c | c c c | c c c | c c c | r}
$\sigma$ & D &$\tau_{s}$ & $\gamma_{s}$ & $\beta_{s}$ & $\tau_{d}$ & $\gamma_{d}$ & $\beta_{d}$ & $\tau_{a}$ &$\gamma_{a}$ & $\beta_{a}$ & $\langle E \rangle$\\
\hline
0.00 & 2 & 1.28(1) & 0.00 & 2.76(2) & 1.50(1) & 0.00 & 1.53(2)  & 1.35(1) & 0.00 & 2.00(2) & 0.554\\
0.01 & 2 &1.36(1) & 0.50(5) & 2.98(2) & 1.70(1) & 1.2(1) & 1.65(2)  & 1.45(1) & 0.69(9) & 2.00(2)  & 0.554\\
0.03 & 2 &1.58(2) & 2.7(1) & 3.78(5) & 2.07(2) & 2.95(9) & 1.89(4)  & 1.88(2) & 4.0(1) & 2.00(3)  & 0.553\\
0.05 & 2 &1.56(2) & 1.79(5) & 3.87(4) & 2.25(2) & 3.85(9) & 2.00(4)  & 2.10(2) & 5.0(1) & 2.00(4)  & 0.551\\
0.10 & 2 &1.54(2) & 0.8(1) & 3.85(5) & 2.20(2) & 2.75(9) & 2.00(4)  & 2.20(2) & 5.00(5) & 2.00(4)  & 0.545\\
0.15 & 2 &1.52(2) & 0.5(1) & 3.96(5) & 2.20(2) & 2.4(1) & 2.00(5)  & 2.21(2) & 4.5(1) & 2.00(5)  & 0.534\\
0.20 & 2 &1.51(2) & 0.35(9) & 4.00(5) & 2.14(2) & 1.8(1) & 2.00(5)  & 2.20(2) & 4.0(1) & 2.00(5)  & 0.521\\
0.23 & 2 &1.50(2) & 0.00 & 4.00(5) & 2.00(2) & 0.7(1) & 2.00(5)  & 2.20(2) & 3.9(1) & 2.00(5)  & 0.512\\
0.25 & 2 &1.50(2) & 0.00 & 4.00(5) & 2.00(2) & 0.7(1) & 2.00(5)  & 2.20(2) & 3.6(1) & 2.00(5)  & 0.505\\
\hline
0.00 & 3 &1.41(2) & 0.00 & 3.40(5) & 1.76(2) & 0.00 & 1.72(5)  & 1.42(2) & 0.00 & 3.00(5)  & 0.525\\
0.05 & 3 &1.43(2) & 0.00 & 3.46(5) & 1.82(2) & 0.35(9) & 1.75(5)  & 1.43(2) & 0.00 & 3.00(5)  & 0.523\\
0.12 & 3 &1.44(2) & 0.00 & 3.60(5) & 1.90(2) & 0.7(1) & 1.84(5)  & 1.45(2) & 0.00 & 3.00(5)  & 0.512\\
0.20 & 3 &1.47(2) & 0.00 & 3.70(5) & 2.00(2) & 0.8(1) & 1.84(5)  & 1.47(2) & 0.00 & 3.00(5)  & 0.490\\
0.23 & 3 &1.49(2) & 0.00 & 3.87(5) & 2.00(2) & 0.8(1) & 1.86(5)  & 1.49(2) & 0.00 & 3.00(5)  & 0.478\\
0.30 & 3 &1.50(2) & 0.00 & 3.90(5) & 2.00(2) & 0.8(1) & 1.89(5)  & 1.49(2) & 0.00 & 3.00(5)  & 0.447

\end{tabular}
\caption{Scaling exponents and average energy of two and three dimensional systems for different values of noise level $\sigma$. The exponents are obtained from plots similar to Fig. (\ref{fig1}--\ref{fig4}). The values of the exponents and their uncertainty are obtained using the method presented in the main frame of the plots. }
\label{table1}
\end{table*}

\begin{table}
\setlength{\tabcolsep}{2pt}
\begin{tabular}{l | c | c c c | c c c| r}
$D$ & $\sigma$ & $\tau_{s}$ & $\gamma_{s}$ & $\beta_{s}$ & $\tau_{d}$ & $\gamma_{d}$ & $\beta_{d}$ & $\langle E \rangle$ \\
\hline
   4 & 0.00  & 1.50(2) & 0.00 & 3.60(5) & 2.00(2) & 0.8(1) & 1.80(5) & 0.512\\
   2 & 0.23  & 1.50(2) & 0.00 & 4.00(5) & 2.00(2) & 0.7(1) & 2.00(5) & 0.512\\
   3 & 0.23  & 1.49(2) & 0.00 & 3.87(5) & 2.00(2) & 0.8(1) & 1.86(5) & 0.478\\
   4 & 0.23  & 1.50(2) & 0.00 & 3.80(5) & 2.00(2) & 0.7(1) & 1.82(5) & 0.455
\end{tabular}
\caption{Scaling exponents and average energy for four dimensional SPZ model in comparison to those of a two, three and four dimensional noisy systems with $\sigma=0.23$. Exponents and uncertainties are obtained similar to that in Table \ref{table1}}
\label{table2}
\end{table}

\section{Concluding remarks}

Motivated by recent experiments on neuronal avalanches we have studied a continuous sandpile model of SOC in the presence of local noise. The model simulates how an active neuron redistributes its load unevenly down the synaptic pathways and suffers noisy transfer of charge at the synaptic junctions, leading to possible activity of post-synaptic neurons. The fact that the average noise at the synaptic junctions is assumed to be zero is related to the global charge conservation in the brain during empirical observation times. We find that if the local noise is on the average conservative ($\langle \eta \rangle=0$) the system remains critical, with well-defined critical exponents, which we obtain from finite-size scaling collapses. More interestingly, we find that as the noise level is increased, such exponents move toward their mean-field values and finally saturate at relatively large values of noise. We looked at both $2D$ and $3D$ systems and found this trend to be true. Such mean-field exponents are exactly what are seen in neuronal avalanche experiments. It is generally believed that such mean-field exponents are due to high connectivity network structure of the brain. Here, we have provided an independent mechanism to produce mean-field results, i.e. that of noisy local dynamics. Mean-field solutions are theoretically obtained where all to all connections are assumed. Numerically, such exponents have been found in various sandpile models with small-world effect, e.g. scale-free networks \cite{GLKK} or small-world networks \cite{AH}. More recently, the existence of two scaling regimes, one of mean-field-like and one of regular-lattice-like have also been reported \cite{HM,BS}. We note that in all such studies definitive mean-field behavior has been observed for the value of shortcut link probability (long range connections) equal or greater than $p\sim 10^{-1}$. Since typically a neuron has $10^{4}$ connections, and \textit{assuming} that a small number of such connections are long-ranged, one may need to study the behavior of such models for $p\ll10^{-1}$, where we do not expect to get mean-field exponents over the entire range of data. We therefore suspect that the observed mean-field exponents in real experiments are results of both small-world structure of neural networks as well as their noisy dynamics \cite{smam}.

Moreover, independent of neuroscience applications of the work presented here, we have addressed an important question. Since it is generally believed that violation of local conservation leads to lack of criticality \cite{foot3}, we have shown that if one breaks local conservation but respects global conservation on the average, one still preserves criticality in such systems. We have established this fact by extensive finite-size scaling (as well as time series) analysis. The exponents of such critical systems depend on the intensity of noise and saturate at their mean-field values for sufficiently strong noise level. We have additionally provided arguments as to why large local noise should lead to high-dimensional mean-field behavior.

Finally, we note that our original SPZ model belongs to what is known as the Manna universality class \cite{SV}. It is generally believed \cite{BB,KNWZ,MDVZ} that stochastic/deterministic as well as isotropic/directed local dynamics are the relevant symmetries that define various universality classes in the sandpile models. Since the addition of noise into our local dynamics does not change any of the relevant symmetry properties (stochastic and global isotropicity) one would still expect the model to remain in the Manna universality class. We on the other hand, observe a gradual change of exponents to their mean-field values. Therefore, our results provide evidence for a critical sandpile model with stochastic undirected dynamics which does not belong to the Manna universality class, requiring us to reconsider the controversial concept of universality in sandpile models \cite{CMV,GJ}. We close by noting that the mean-field behavior observed in our model is due to an \emph{effective} range of \lq\lq interaction" which increases with increasing noise, which underlies the mean-field universality class. The gradual change of the exponents may be interpreted as a crossover effect which would take various sandpile models to mean-field behavior, as the range of interaction gradually increases from short-range to long-range behavior, with increasing noise. It would be interesting to investigate whether such behavior is generic to all sandpile (or SOC) models.

\begin{acknowledgments}Support of Shiraz University Research Council is kindly acknowledged. We would like to thank two anonymous referees whose constructive criticisms help improve our paper.
\end{acknowledgments}

\bibliographystyle{apsrev}

\begin{thebibliography}{99}

\bibitem{BTW1}  P. Bak, C. Tang, and K. Wiesenfeld, Phys. Rev. Lett. \textbf{59}, 381 (1987).

\bibitem{BTW2}  P. Bak, C. Tang, and K. Wiesenfeld, Phys. Rev. A \textbf{38}, 364 (1988).

\bibitem{B}     P. Bak, \textit{How Nature Works: the Science of Self-Organized Criticality} (Springer-Verlag New York, 1996).

\bibitem{CM}    K.Christensen, N. R. Moloney, \textit{Complexity and Criticality} (Imperial college press London, 2005).

\bibitem{P}     G. Pruessner, \textit{Self-Organised Criticality: Theory, Models and Characterisation} (Cambridge University Press New York, 2012).

\bibitem{foot1}  There are some exceptions to this generel rule where models with finite drive can be mapped to sandpile models, see for example, [H. Takayasu, Phys. Rev. Lett. \textbf{63}, 2563 (1989)] and \cite{D}.

\bibitem{HK1}   T. Hwa, and M. Kardar, Phys. Rev. Lett. \textbf{62},16 (1989).

\bibitem{HK2}   T. Hwa, and M. Kardar, Phys. Rev. A \textbf{45},7002 (1992).

\bibitem{GLS}   G. Grinstein, D. -H. Lee, and S. Sachdev, Phys. Rev. Lett. \textbf{64},16 (1990).

\bibitem{OFC}   Z. Olami, H. J. S. Feder, and K. Christensen, Phys. Rev. Lett. \textbf{68},1244 (1992).

\bibitem{CO}   K. Christensen, and Z. Olami, Phys. Rev. A \textbf{46}, 1829 (1992).

\bibitem{DS}   B. Drossel, and F. Schwabl, Phys. Rev. Lett. \textbf{69}, 1629 (1992).

\bibitem{BM}    J. A. Bonachela, and M. A. Mu\~{n}oz, J. Stat. Phys. P09009 (2009).

\bibitem{BP1}    J. M. Beggs, and D. Plenz, J. Neurosci. \textbf{23}, 11167 (2003).

\bibitem{BP2}    J. M. Beggs, and D. Plenz, J. Neurosci. \textbf{24}, 5216 (2004).

\bibitem{FIBSLD} N. Friedman, S. Ito, B. A. W. Brinkman, M. Shimono, R. E. Lee DeVille, K. A. Dahmen, J. M. Beggs, and T. C. Butler, Phys. Rev. Lett. \textbf{108}, 208102 (2012).

\bibitem{PTLNCP} T. Petermann, T. C. Thiagarajan, M. A. Lebedev, M. A. L. Nicolelis, D. R. Chialvo, and D. Plenz, PNAS, \textbf{106}, 37 (2009).

\bibitem{TBFC} E. Tagliazucchi, P. Balenzuela, D. Fraiman, and D. R. Chialvo, Front. Physiol. \textbf{3}, 15 (2012).

\bibitem{SACHHSCBP} O. Shriki, J. Alstott, F. Carver, T. Holroyd, R. N. A. Henson, M.L. Smith, R. Coppola, E. Bullmore, and D. Plenz, J. Neurosci. \textbf{33}, 7079 (2013).

\bibitem{HTBC}  A. Haimovici, E. Tagliazucchi, P. Balenzuela, and D. R. Chialvo, Phys. Rev. Lett. \textbf{110}, 178101 (2013).

\bibitem{C}     D. Chialvo, Nature Phys. \textbf{6}, 744 (2010).

\bibitem{ABBOTT} P. Dayan, and L. F. Abbott, \textit{Theoretical Neuroscience: Computational and Mathematical Modeling of Neural Systems}, (MIT press, 2005).

\bibitem{RD}    E. T. Rolls, G. Deco, \textit{The Noisy Brain: Stochastic Dynamics as a Principle of Brain Function}, (Oxford University Press Oxford, 2010).

\bibitem{AM}    A. Abdolvand, and A. Montakhab, Eur. Phys. J. B \textbf{76}, 21 (2010).

\bibitem{Z}     Y. -C. Zhang, Phys. Rev. Lett. \textbf{63}, 470 (1989).

\bibitem{VP}     V. Privman, \textit{Finite Size Scaling and Numerical Simulation of Statistical Systems} (World Scientific Singapore, 1990).

\bibitem{SV}    R. Pastor-Satorras, and A. Vespignani, Eur. Phys. J. B \textbf{18}, 197-200 (2000).

\bibitem{Manna} S.S. Manna, J. Phys. A \textbf{24}, L363 (1991).


\bibitem{D}     D. Dhar, Physica A \textbf{263}, 4 (1999).

\bibitem{D1}    D. Dhar, Physica A \textbf{369}, 29 (2006).

\bibitem{MSB}   O. Malcai, Y. Shilo, and O. Biham, Phys. Rev. E \textbf{73}, 056125 (2006).

\bibitem{Dic}   R. Dickman, Phys. Rev. E \textbf{73}, 036131 (2006).


\bibitem{A}     P. Alstr{\o}m, Phys. Rev. A \textbf{38},9 (1988).

\bibitem{ZLS}   S. Zapperi, K. B. Lauritsen, and H. E. Stanley, Phys. Rev. Lett. \textbf{75}, 4071 (1995).

\bibitem{DC}    R. Dickman, and J. M. M. Campelo, Phys. Rev. E \textbf{67}, 066111 (2003).

\bibitem{GLKK}  K. -I. Goh, D. -S. Lee, B. Kahng, and D. Kim, Phys. Rev. Lett. \textbf{91},14 (2003).

\bibitem{LU}    S. L\"{u}beck, Phys. Rev. E \textbf{58}, 2957 (1998).

\bibitem{W}     K. G. Wilson, and J. Kogut, Phys. Rep. \textbf{12C}, 75 (1974).

\bibitem{MC}   A. Montakhab, and J. M. Carlson, Phys. Rev. E \textbf{58}, 5608 (1998).

\bibitem{foot2}  Not only one is limited by relatively small range of $L$, one typically bins (or averages) the obtained statistics over a given range in order to smooth out the resulting distribution functions.

\bibitem{MSP}   E. Martin, A. Shreim, and M. Paczuski, Phys. Rev. E \textbf{81}, 016109 (2010).

\bibitem{MM}    A. Montakhab, and S. Mohammadpour, (in preparation).

\bibitem{AH}   L. de Arcangelis, and H. J. Herrmann, Physica A \textbf{308}, 545 (2002).

\bibitem{HM}   M. Hoore, and S. Moghimi-Araghi, J. Phys. A \textbf{46}, 195001 (2013).

\bibitem{BS}   H. Bhaumik, and S. B. Santra, Phys. Rev. E \textbf{88}, 062817 (2013).

\bibitem{smam} S. A. Moosavi, A. Montakhab, (in preparation).

\bibitem{foot3} The self-organized directed percolation model which does not include a conservation law (nor any dissipation) is an important exception, see [P. Grassberger, and Yi-Cheng Zhang, Physica A \textbf{224}, 169 (1996).]


\bibitem{BB}   A. Ben-Hur, and O. Biham, Phys. Rev. E \textbf{53}, R1317 (1996).

\bibitem{KNWZ} L. P. Kadanoff, S.R. Nagel, L. Wu, and S. Zhou, Phys. Rev. A \textbf{39}, 6524 (1989).

\bibitem{MDVZ}  M. A. Mu\~{n}oz, R. Dickman, A. Vespignani, and S. Zapperi, Phys. Rev. E \textbf{59}, 6175 (1999).

\bibitem{CMV}   A. Chessa, E. Marinari, and A. Vespignani, Phys. Rev. Lett. \textbf{80}, 4217 (1998).


\bibitem{GJ}    A. Giometto, and H. J. Jensen, Phys. Rev. E \textbf{85}, 011128 (2012).



\end{thebibliography}

\end{document}